\documentclass[aps,prb,reprint,groupedaddress,footinbib,citeautoscript]{revtex4-1}

\usepackage{graphicx}
\usepackage[utf8]{inputenc}
\usepackage{csquotes}
\usepackage[american]{babel}
\usepackage[T1]{fontenc}
\usepackage{enumerate}
\usepackage{mdwlist}
\usepackage[activate=normal]{pdfcprot}
\usepackage{bbding}
\usepackage{color}
\usepackage{amssymb}
\usepackage{amsmath}
\usepackage{amsfonts}
\usepackage{mathrsfs}

\usepackage{bm}
\usepackage{dcolumn}
\usepackage{color}
\usepackage[colorlinks=true,citecolor=blue]{hyperref}
\hypersetup{colorlinks=true,citecolor=blue,linkcolor=red,urlcolor=blue}

\begin{document}

\title{Gate controllable spin pumping in graphene via rotating magnetization}

\author{Mojtaba A. Rahimi}
\affiliation{Department of Physics, Institute for Advanced Studies in Basic
Sciences (IASBS), Zanjan 45137-66731, Iran}

\author{Ali G. Moghaddam}
\affiliation{Department of Physics, Institute for Advanced Studies in Basic
Sciences (IASBS), Zanjan 45137-66731, Iran}

\date{\today}

\begin{abstract}
We investigate pure spin pumping in graphene by imposing a
ferromagnet (F) with rotating magnetization on top of it.
Using the generalized scattering approach for adiabatic spin pumping, we obtain the  spin current pumped through magnetic graphene to a neighboring  normal (N) region. The spin current can be easily controlled by gate voltages and under certain conditions, becomes sufficiently large to be measurable in current experimental setups. In fact  it reaches a maximum value when one of the spins are completely filtered due to the vanishing density of states of the corresponding spin species in the ferromagnetic part. 
Considering an N|F|N structure with a finite ferromagnetic region, it is found that in contrast to the metallic ferromagnetic materials the transverse spin coherence length can be comparable to the length of F denoted by $L$. Subsequently, due to the quantum interferences, the spin current becomes oscillatory function of $JL/\hbar v_F$ in which $J$ is the spin splitting inside F. Finally we show, originated from the controllability of pumped spin into two different normal sides, a profound spin battery effect can be seen in the hybrid N|F|N device. 
\end{abstract}

\maketitle

\section{Introduction}
One of the important goals and challenges of spintronics is the generation and sensing of spin current.
During the last decades, several ways have been proposed and implemented to generate spin current such as electrical spin injection \cite{Ohno1999,Rashba2000,Jedema2001,Zutic2004}, optical methods \cite{Taniyama2011}, spin hall effect \cite{Murakami2003, Sinova2004, Kato2004, Wunderlich2005,Saitoh2006}, etc.
In most of these approaches, spin injection is associated with a dissipative charge current which strongly affects the functionality of any spintronic device. Nevertheless 
the demand of pure spin generation (without charge current) is increased a lot by promising opportunities introduced by effects like spin wave motion in magnetic insulators \cite{Kajiwara2010}, and spin Seebeck effects in the so-called spin caloritronic systems \cite{Uchida2008,Bauer2012}. Intriguingly the so called spin pumping provides a unique possibility of generating dissipationless spin currents without any external bias in metallic systems. 
Considering a precessing ferromagnet in an externally applied magnetic field, a pure spin current will be pumped into the adjacent normal metal \cite{Tserkovnyak2002,Lenz2004,Mosendz2010}.
Spin pumping in this way originates from the lose of angular momentum inside the ferromagnet and in this regard, the spin pumping by precessing ferromagnet can be assumed as the reverse process of the so-called spin transfer torque effect (STT) \cite{Brataas2006,brataas2011}.An immediate result of this reciprocity relation between spin pumping and STT is the fact that the spin pumping is accompanied by the enhancement of the so-called Gilbert damping \cite{Tserkovnyak2002a,Zwierzycki2005}.
In recent years various applications in nanoelectronics have been suggested for spin pumping, in which among them the spin battery effect has received a considerable attention \cite{Brataas2008}.
\par
During last years, graphene, as a leading material among atomic monolayers has received a tremendous amount of interest mostly due to its very special electronic structure described by the massless Dirac
model \cite{Geim2007,neto2009}. At the same time a diverse variety of applications in chemistry, electronics, optics and other industries have been suggested for it \cite{Gilje2007,Geim2009,Novoselov2012}. Intriguingly, beside other applications recently graphene has found promising applications in spintronics \cite{Fabian2014}, due to the very long spin relaxation length  \cite{Tombros2007,Jozsa2008,Tombros2008,Zomer2012,
wees2012}.
On the other hand a variety of methods have been proposed to
induce magnetism or spin splitting in graphene.
In practice one may put an insulating ferromagnetic on top of graphene \cite{Hernando2008} or, alternatively, add magnetic impurities and adatoms to generate spin imbalance \cite{eelbo}. In addition it is believed that under certain conditions intrinsic magnetism may be generated due the edges or defects in it
(see Ref. \onlinecite{yazyev} for magnetism in graphene). Interestingly,
the spin splitting in graphene if becomes large enough and comparable with Fermi energy of the system with respect to the Dirac point, leads to interesting effects including spin focusing \cite{moghaddam2010} among other spin dependent transport features \cite{moghaddam2008,zare2013}.
\par
In recent years, both electrical spin injection \cite{Tombros2007,Jozsa2008,Tombros2008} and adiabatic spin pumping by exerting an AC gate voltage \cite{Zhang2011} have been theoretically and experimentally studied in graphene. All these studies which take the advantage of ferromagnetic proximity effect, are associated with charge currents. Subsequently the efficiency of the pumped spin current is relatively small due to the conductance mismatch problem.
To overcome this issue it has been suggested to use tunnel barriers resulting in higher efficiencies \cite{Han2012}. Very recently, the dynamical pure spin pumping approach (which is free of conductance mismatch) to generate pure spin currents in monolayer graphene has attracted significant attention \cite{Patra2012,Tang2013,Singh2013}.
\par
In this paper the possibility of pure spin pumping in graphene is investigated by considering a ferromagnetic insulator on top of it with rotating magnetization which induces an exchange splitting \cite{Semenov2008}. We employ the generalized form of Brouwer’s formula for adiabatic parametric pumping \cite{Brouwer}, provided by Tserkovnyak \emph{et al.} \cite{Tserkovnyak2002} in order to include spin degree of freedom for our calculations. First, the effect of electron concentration which can be varied by gate voltages, on pure spin pumping in an F|N structure is investigated. We find that the resulting spin current can have a considerable value which is measurable in the currently available experiments (for instance see Ref. \onlinecite{Patra2012,Tang2013,Singh2013}
which measure enhanced Gilbert damping). 
The spin pumping is very efficient when $\mu_F=\pm J$ ($\mu_{N,F}$ are the chemical potentials of N an F parts, respectively) and becomes more profound by increasing $\mu_N$. This enhancement comes from the fact that up and down spin species pass through the Dirac point when $|\mu_F|= J$, respectively.
We also appraise an N|F|N structure in order to explore the effect of length of ferromagnetic region on the spin pumping. In contrast to the metallic ferromagnets \cite{Brataas2006}, here the transverse spin coherence length $d_F=\pi/\left(k_{F\uparrow}-k_{F\downarrow}\right)$ [where $k_{F\uparrow}(k_{F\downarrow})$ is Fermi wave vector for spin-up (down)] is found to be comparable to the length of ferromagnetic region and is tunable by the exchange splitting $J$. Since we assume the ballistic transport regime the successive scatterings from two N/F interfaces results in the quantum interference effects which leads to oscillations of the spin pumping with both length of F and the spin splitting. Unlike the conventional metallic systems the tunability of doping in graphene provides the unique possibility of having very different spin currents injected into two normal leads by varying the corresponding gate voltages. The difference could be very large when one of the leads has very low charge carriers density and the spin splitting lies in certain ranges. Therefore the N|F|N structure can work as a spin battery with controllable polarization and power.
\section{The model and basic formalism}
The prototype setup for spin pumping consists of a ferromagnetic region sandwiched between two normal parts. All transport properties of such a system in the ballistic regime can be fully described versus its scattering matrix,
\begin{equation}
\hat{S}=\left(\begin{matrix}
\hat{r} &\hat{t'}\cr
\hat{t} &\hat{r'}
\end{matrix}\right)
\end{equation}
in which $\hat{r}$ and $\hat{t}$ are reflection and transmission matrices for the incoming electrons from left and $\hat{r'}$ and $\hat{t'}$ are corresponding matrices for the electrons incident from right. In the absence of voltage bias and magnetization dynamics, there will be no net charge and spin currents.
When the magnetization vector ${\bf M}$ starts to precess then a pure  spin current without any charge pumping will be generated. The pumped spin current contains both \emph{d.c.} and \emph{a.c.} components which in the adiabatic limit are proportional to the precession frequency $\omega$. Here, we consider the magnetization vector rotates around $y$-axis in the $xz$-plane. Then the \emph{d.c.} part of the pumped spin current in the $y$-direction (perpendicular to the magnetization precession plane) through the normal region is given by \cite{Tserkovnyak2002},
\begin{equation}\label{Is2}
{\bf I}_{s}=\frac{\hbar \omega}{4\pi}{\rm Re}(g^{\uparrow\downarrow}-t^{\uparrow\downarrow}) \hat{\bf y}.
\end{equation}
where $g^{\uparrow\downarrow}=N-{\rm Tr}(\hat{r}^{\uparrow}\hat{r}^{{\downarrow}\dag})$ is the mixing conductance and $t^{\uparrow\downarrow}={\rm Tr}(\hat{t'}^{\uparrow}\hat{t'}^{{\downarrow}\dag})$. The matrices $\hat{r}^{s}$ and $\hat{t'}^{s}$ describe the reflection and transmission of spin $s$ electrons which are scattered to the left part of the scattering region.
\par
The low energy dispersion in the ferromagnetic monolayer graphene is describe by the Dirac Hamiltonian
\begin{equation}\label{Hamiltoni}
\textrm{H}=\hbar v_F \left(\textbf{k} \cdot\hat{\bm \sigma}  \otimes \hat{s}_0   \right)-J\hat{\sigma}_0 \otimes \hat{s}_z +\mu_F
\end{equation}
Here $\textbf{k}=(k_x,k_y)$ is the momentum of quasiparticles,$v_F\approx 10^6 m/s$ is the Fermi velocity. Matrices $\hat{\sigma}_{i}$ and $\hat{s}_{i}$ ($i=0,1,2$) are the Pauli matrices in the so-called pseudospin and spin subspaces, respectively. Moreover $J$ denotes the strength of exchange field induced by a ferromagnetic insulator and $\mu_F$ is the chemical potential which can be varied by gate voltages.
\par
\begin{figure}[t]

\includegraphics[width=0.9 \linewidth]{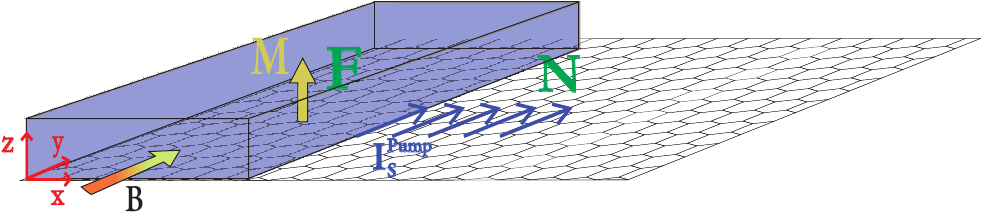}
\put (-200.5, 50){\makebox[0.01\linewidth][r]{(a) }}\vspace{0.5 cm}
\includegraphics[width=0.9 \linewidth]{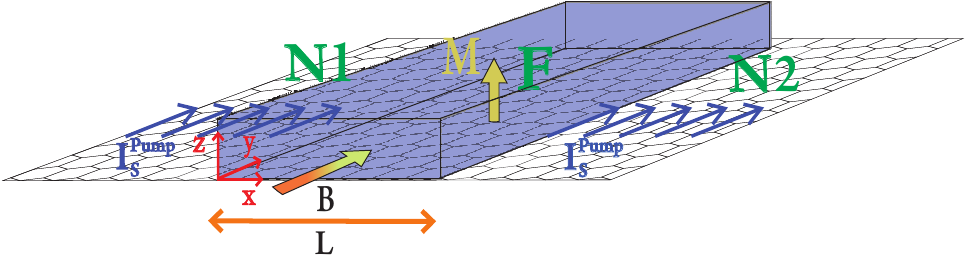}
\put (-200.5, 50){\makebox[0.01\linewidth][r]{(b) }}
\caption{\label{fig1} Schematic graphene based spin pumping devices: (a) F|N, (b) N|F|N structure. The magnetization rotates around static magnetic field which is in $y$ direction. The blue arrows show the pumped spin current.} 
\end{figure}

We consider two different setups as shown in Fig. \ref{fig1}. The first device
contains only one F/N interface and we will study the pumped spin current from magnetic region to normal region. But the second one consists of a finite magnetic region joined to two normal parts. So in this case the spin current will be pumped into two different N regions. In both cases by exerting a large static magnetic field in the $y$ direction beside an oscillating \emph{rf} part in the perpendicular plane, the magnetization vector of the ferromagnetic insulator starts to precess. Then a precessing exchange field will be induced in the monolayer graphene where the exchange strength is independent of the rotation angle \cite{Semenov2008}. So the rotating magnetization solely lift the degeneracy of up and down spins.
\subsection{ N|F structure}
In order to obtain the pumped spin current into the normal region for the N|F structure as in Fig.\ref{fig1}(a), we need the scattering properties of the interface.  
Assuming an electron incident from N to F part, the scattering wave function in two regions reads,
\begin{eqnarray}
\nonumber \psi_N^{s} &=& \frac{1}{\sqrt{\cos{\alpha}}}\left(
                                                     \begin{array}{c}
                                                       \chi_{s} {\rm sgn}(\mu_N) e^{-i\alpha} \\
                                                       \chi_{s} \\
                                                     \end{array}
                                                   \right) e^{ik_x x} \\
&+&\frac{ r^{s}}{\sqrt{\cos{\alpha}}}\left(                                                \begin{array}{c}
  -\chi_{s} {\rm sgn}(\mu_N) e^{i\alpha} \\
    \chi_{s} \\
    \end{array}
 \right) e^{-ik_x x}
  \\
 \nonumber \psi_F^{s} &=& \frac{ t^{s}}{\sqrt{\cos{\beta_{s}}}}\left(
                                                     \begin{array}{c}
                                                       \chi_{s} {\rm sgn}(\mu_F-s J) e^{-i\beta_{s}} \\
                                                       \chi_{s} \\
                                                     \end{array}
                                                   \right) e^{ik'_{xs } x}
\end{eqnarray}
with the spinors $\chi^\dag_{\uparrow}=\left(1,0\right)$ and $\chi^\dag_{\downarrow}=\left(0,1\right)$.
The incident and transmitted electron angles are $\alpha=\arctan(k_y/k_x)$ and $\beta_{\sigma}=\arctan(k_y/k'_{x s })$, respectively, in which $k_x=\sqrt{(\mu_N/\hbar v_F)^2- k_y^2}$
and $k'_{x s }=\sqrt{(\mu_F - s  J)^2/(\hbar v_F)^2- k_y^2}$ indicate the longitudinal momentum of excitations inside N and F parts. 
The transmission and reflection coefficients can be calculated by writing the wave functions in N and F regions and imposing the boundary conditions on them.
Since only the reflection coefficients insert the mixing conductance we only write the result for $r^{ s }$. Depending on the transverse momentum $k_y$ of the incident electron, the corresponding state inside the F region can be propagating (if $|k_y|<(\mu_F+ s  J)/\hbar v_F$) or evanescent (if $|k_y|>(\mu_F+ s  J)/\hbar v_F$). For the propagating case the reflection amplitude has an absolute value of less than one given by,
\begin{equation}
r^{ s }=\frac{-\text{sgn}(\mu_F - s  J) +\text{sgn}(\mu_N) e^{-i (\alpha -\beta_{ s } )}}{\text{sgn}(\mu_F - s  J) +\text{sgn}(\mu_N) e^{i (\alpha +\beta_{ s } )}}.
\end{equation}
But if the state becomes evanescent inside F then the electron is completely reflected with, 
\begin{eqnarray}
&&r^{ s }=\frac{-i \,\text{sgn}(\mu_F - s  J) e^{ -\text{sgn}(k_y)\phi_{ s }}-e^{-i\, \alpha \, \text{sgn}(\mu_N )}}{i\, \text{sgn}(\mu_F - s  J) e^{ -\text{sgn}(k_y)\phi_{ s }}-e^{i\, \alpha \, \text{sgn}(\mu_N )}}
\\
&&\phi_{ s } =\cosh ^{-1}\left(\frac{ \hbar v_F k_y}{\left| \mu_F - s  J\right| }\right).
\end{eqnarray}
For the pumped current we need to find the transmission amplitudes from F to N as well. This can be done by assuming scattering state with electron incident from F to the interface which reads,

\begin{equation}
 {t'}^ s  = \frac{ \sqrt{ \cos \alpha} \left(1+e^{2 i \beta_{ s } }\right) \text{sgn}(\mu_F -{ s }J)}{\sqrt{\cos \beta_{ s }} \left(\text{sgn}(\mu_F - s  J)+e^{i (\alpha +\beta_{ s } )} \text{sgn}(\mu_N )\right)}
\end{equation}

\subsection{ N|F|N structure}
Unlike conventional ferromagnetic metals in which the magnetic coherence length is on the order of atomic distances, in the system under study it is possible to have large $d_F$. For instance assuming spin splitting $J$ varying from $10$ to $100$ meV, the corresponding transverse spin coherence length $d_F= \pi\hbar v_F/J$ varies approximately from $50$ to $5$nm. Therefore we may expect strong quantum interference effects when the length of the finite magnetic region $L$ is comparable with $d_F$. Similar to the previous setup here we can write down the scattering state in the most general form as below,
\begin{eqnarray}
 \psi_{N1}^{ s } &=& \frac{z_L}{\sqrt{\cos{\alpha}}}\left(
                                                     \begin{array}{c}
                                                       \chi_{ s } {\rm sgn}(\mu_N) e^{-i\alpha} \\
                                                       \chi_{ s } \\
                                                     \end{array}
                                                   \right) e^{ik_x x} \\\nonumber&+&\frac{z_L r^{ s }+z_R t'^{ s }}{\sqrt{\cos{\alpha}}}\left(
                                                     \begin{array}{c}
                                                       -\chi_{ s } {\rm sgn}(\mu_N) e^{i\alpha} \\
                                                       \chi_{ s } \\
                                                     \end{array}
                                                   \right) e^{-ik_x x}
  \\
 \nonumber \psi_F^{ s } &=& \frac{A}{\sqrt{\cos{\beta_{ s }}}}\left(
                                                     \begin{array}{c}
                                                       - \chi_{ s } {\rm sgn}(\mu_F- s  J) e^{i\beta_{ s }} \\
                                                       \chi_{ s } \\
                                                     \end{array}
                                                   \right) e^{-ik_x x} \\\nonumber&+&\frac{B}{\sqrt{\cos{\beta_{ s }}}}\left(
                                                     \begin{array}{c}
                                                       \chi_{ s } {\rm sgn}(\mu_F -  s  J) e^{-i\beta_{ s }} \\
                                                       \chi_{ s } \\
                                                     \end{array}
                                                   \right) e^{-ik_x x}
 \\
\nonumber \psi_{N2}^{ s } &=& \frac{z_R r'^{ s }+z_L t^{ s }}{\sqrt{\cos{\alpha}}}\left(
                                                     \begin{array}{c}
                                                       \chi_{ s } {\rm sgn}(\mu_N) e^{-i\alpha} \\
                                                       \chi_{ s } \\
                                                     \end{array}
                                                   \right) e^{ik_x x} \\\nonumber&+&\frac{z_R}{\sqrt{\cos{\alpha}}}\left(
                                                     \begin{array}{c}
                                                       -\chi_{ s } {\rm sgn}(\mu_N) e^{i\alpha} \\
                                                       \chi_{ s } \\
                                                     \end{array}
                                                   \right) e^{-ik_x x}
\end{eqnarray}
where for the case of electron incident from left (right) N region, $z_L=1$, $z_R=0$ ($z_L=0$, $z_R=1$). Then we obtain the needed scattering coefficients $r^{ s }$ and ${t'}^{ s }$ for calculating the spin current pumped to left normal region.  
\begin{figure}[tp]
 \includegraphics[width=0.9 \linewidth]{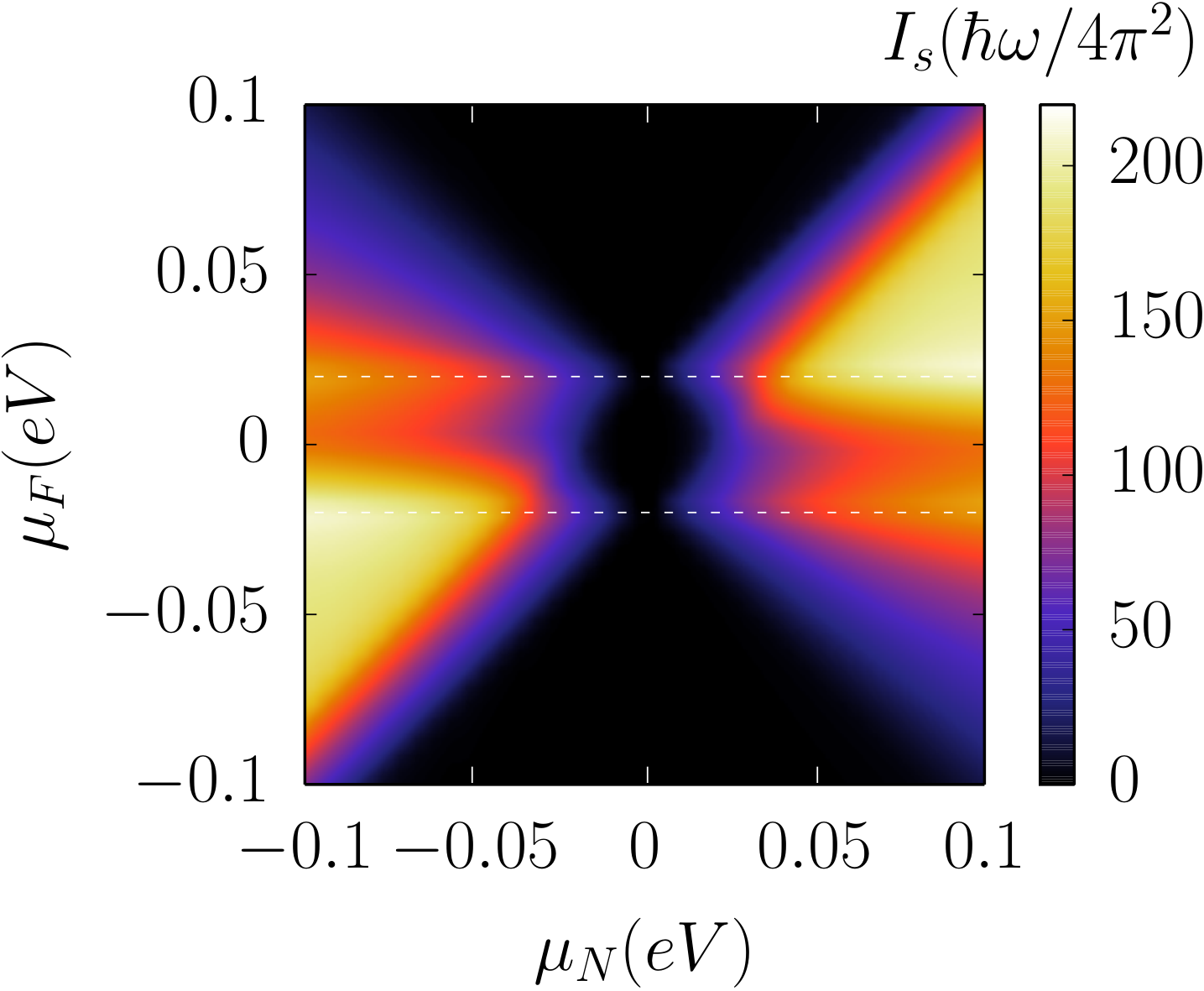}
 \caption{\label{fig2} Pumped spin current into the normal region as a function of gate voltages applied to the N ($\mu_N$) and F ($\mu_F$) regions. The strength of spin splitting is $J=20 meV$ and we have considered a graphene sheet of width $W=2 \mu m$. }
\end{figure}
\section{Numerical Results and Discussion}
\subsection{N|F structure}
First the pumped spin current for the N|F system is calculated numerically as a function of doping of the N and F regions ($\mu_N$ and $\mu_F$).
Figure \ref{fig2} shows the spin pumping variation with both gate voltages for the spin splitting $J=20$meV where we see that inside the region given by $|\mu_N|<||\mu_F|-J|$ the spin pumping is strongly suppressed. This region corresponds to the situation where the Fermi levels of both spin subbands measured from Dirac point are higher than normal region's Fermi level. Therefore any incident electron from normal region has a finite probability of transmission inside the F part irrespective of its spin. Accordingly the spin splitting does not play an important role and the spin pumping is almost absent.
\begin{figure}[tp]
  \includegraphics[width=0.8 \linewidth]{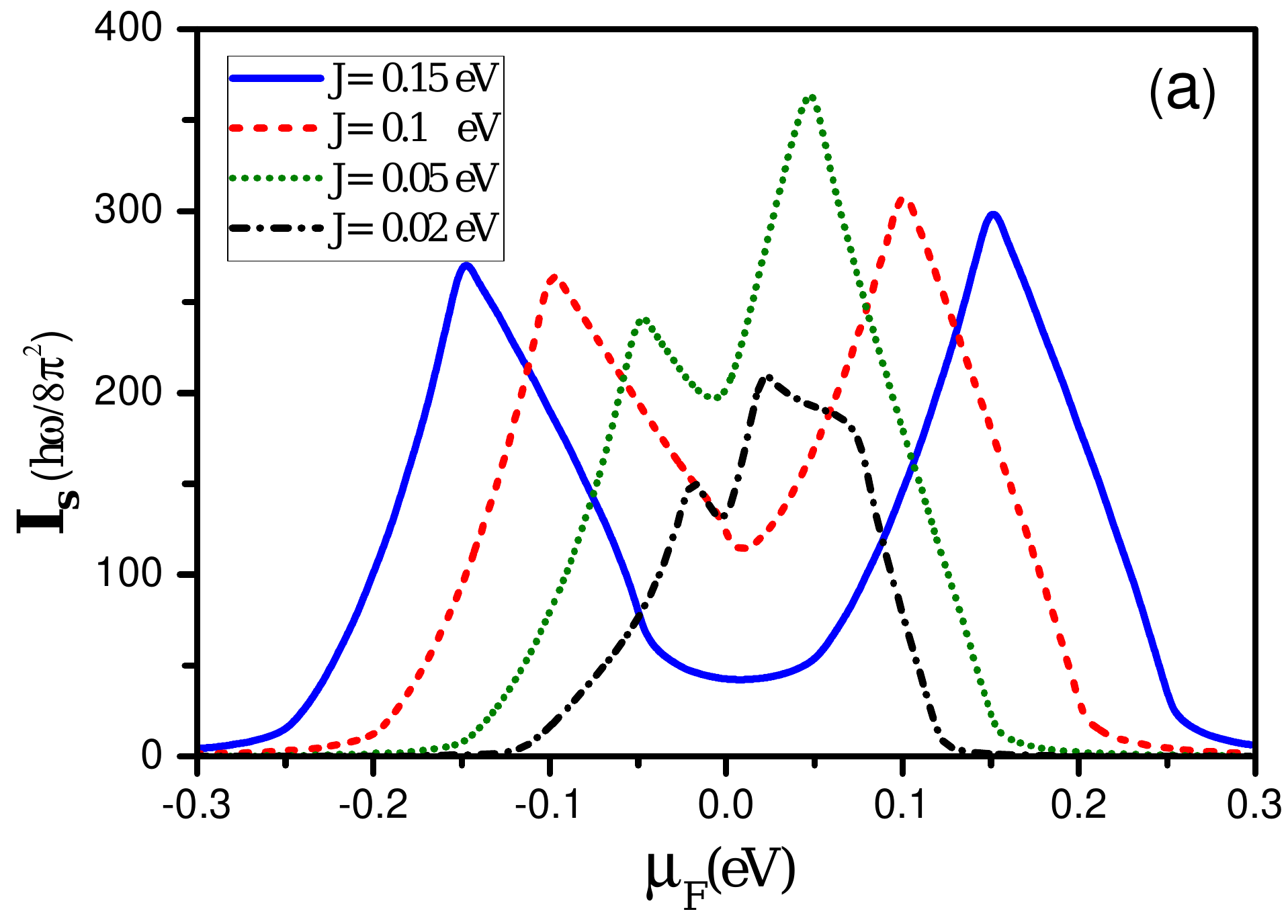}
 \includegraphics[width=0.8 \linewidth]{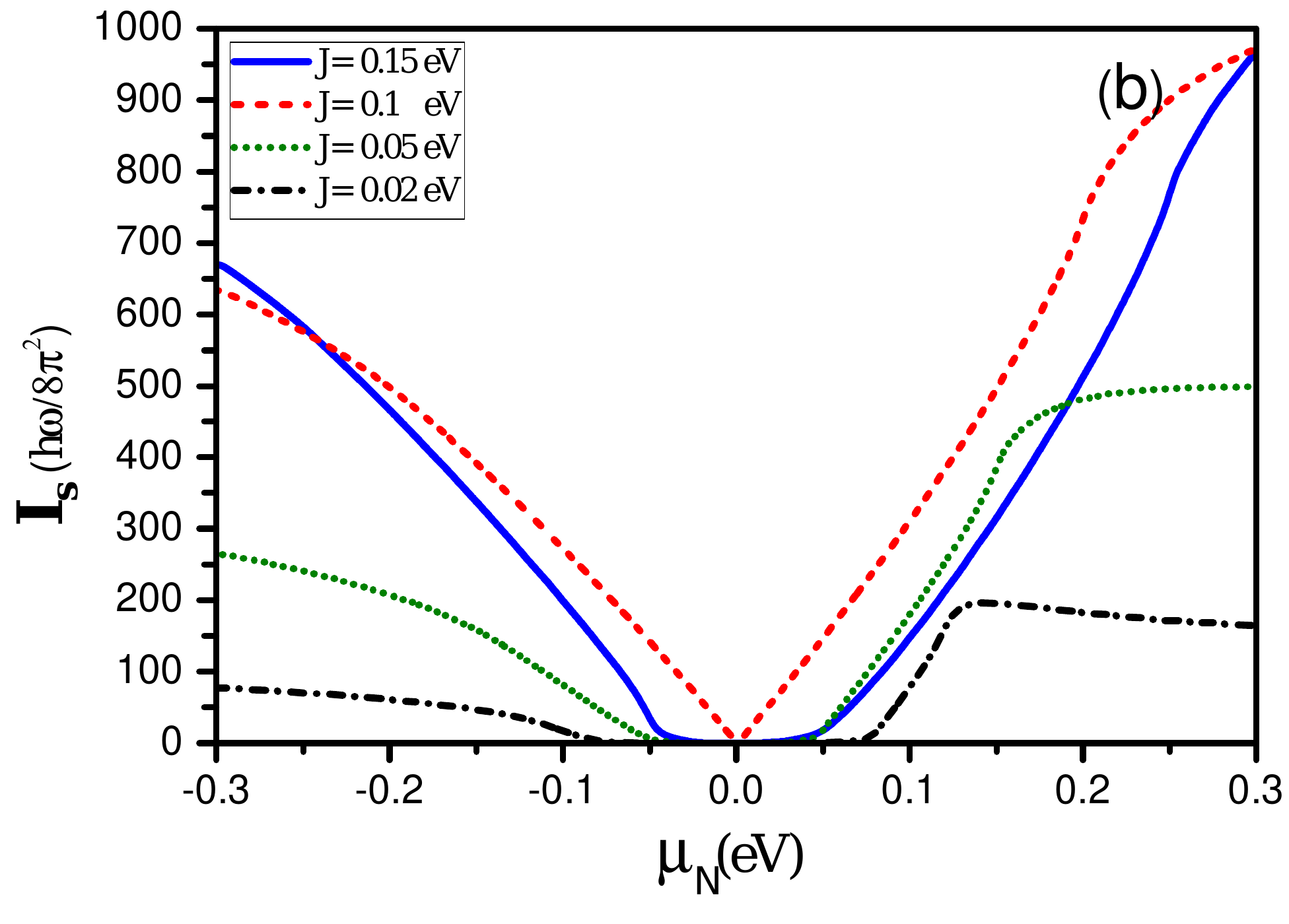}
  \caption{\label{fig3}
Spin current variation with (a) $\mu_F$ and (b) $\mu_N$ for various spin splittings. I each plot the doping of the other lead is fixed to (a) $\mu_N= 0.1$eV and (b) $\mu_F= 0.1$eV. The width of the whole graphene sheet is $W=2 \mu m$.   }
\end{figure}
\par
It is clear that due to the electron-hole symmetry
of the graphene band structure, any transport property including the spin pumping must not change under simultaneous substitutions $\mu_N \to -\mu_N$ and $\mu_F \to -\mu_F$. Based on this and for the sake of clarity we focus only on the situations with $\mu_F > 0$ for the rest of our discussion. Then the spin pumping is finite when the Fermi level inside the normal region becomes larger than minority spin Fermi level inside F ($|\mu_N|>|\mu_F-J|$). The origin of spin pumping in this regime can be related to the fact that only one spin species can participate in the transport, a phenomenon which we call it \emph{spin filtering} at the N|F interface. This effect occurs for the range of modes $|\mu_F-J|<\hbar v_F |k_y|<|\mu_N|$ in which there is no spin down state inside F region. So the down spins of such modes can only be reflected to the normal side ($|r_{\downarrow}|=1$) and subsequently each of them contributes at least an amount proportional to $1-|r_{\uparrow}|)$ in the pumped spin current. It is worth to note that the absence of spin filtering when $|\mu_N|<|\mu_F-J|$ can explain the suppression of spin pumping for that regime. The spin pumping reaches maximum values for $\mu_F=J$ when the Fermi level of minority (down) spins lies in the Dirac point. Then only majority spins can be transmitted to the ferromagnetic part and spin filtering occurs for the whole range of incoming electrons from N region ($\hbar v_F |k_y|<|\mu_N|$). We should mention that since the doping $\mu_F$ can be varied by applying a gate voltage, the amount of pumped spin can be controlled electronically to reach the maximum spin current. 
\begin{figure}[tp]
\includegraphics[width=0.9 \linewidth]{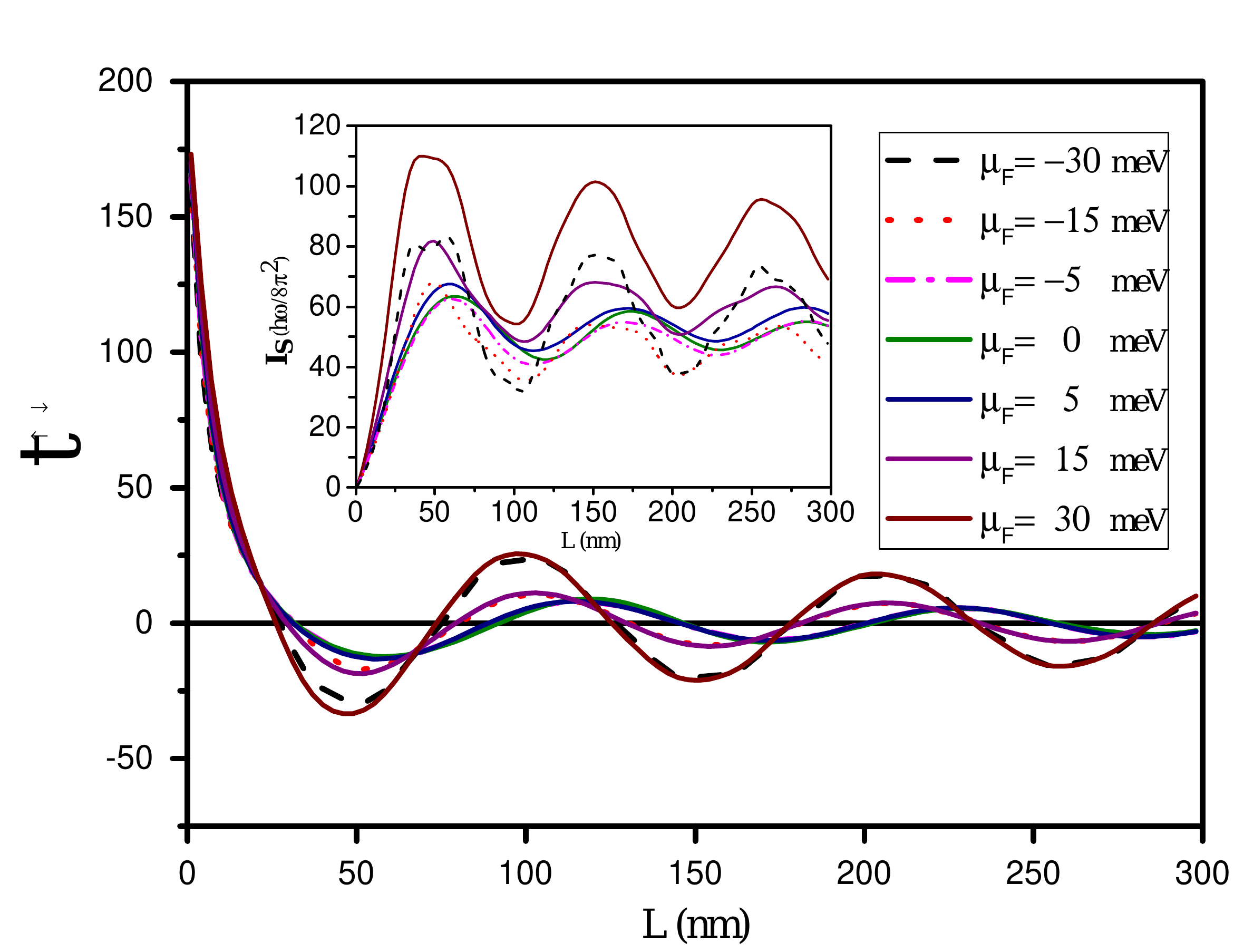}
\caption{\label{fig4}
Illustration of mixing transmission $t^{\uparrow\downarrow}$ as a function of length of the ferromagnetic region for different amounts of Fermi energy (gate voltage). It has a finite considerable value up to the length of the order of $\mu m$ and oscillates over the magnetic coherence length. Inset shows pumped spin current $I_{s,N1}^{\rm pump}$ versus the length of F region. Other parameters are chosen as follow: $\mu_{N1}=\mu_{N2}= 0.1 $eV, $W= 1 \mu m$ and $J= 10$meV }
\end{figure}
\par
In order to see the effect of spin splitting $J$ more clearly in the spin pumping, Fig. \ref{fig3} shows the variation of $I_s$ with normal and ferromagnetic regions doping for various $J$.
It is obvious from these plots that the maximum spin pumping occurs for $|\mu_F|= J$ where one of the spins are completely blocked since the Fermi level passes from Dirac point of the corresponding spin subband. But as a function of normal electrode doping $\mu_N$ pumped spin current increases almost linearly due to the increase in the density of states (DOS). In addition we see an asymmetry with respect to positive and negative dopings of N or F region when the doping inside the other part is fixed. This originates from the fact that spin minority (with smaller DOS) electrons have to pass a barrier of either $n-n$ ($p-p$) or $n-p$ type. This can be seen from Fig. \ref{fig2} as well where in general spin pumping is stronger when the type of carriers in both sides (N and F) is of the same type ($n$ or $p$).
\subsection{N|F|N structure}
Now we turn to the N|F|N device with a finite-length magnetic region. As we mentioned before in this configuration the quantum interference can take place which results in the oscillatory behavior of the transport phenomena with the length of F region scaled by the magnetic coherence length $d_F$. In fact both mixing conductance and mixing transmission will oscillate as a function of $\sim J L/\hbar v_F$. Figure \ref{fig4} shows the variation of $t^{\uparrow\downarrow}$ with the F region length $L$ which indicates the oscillatory behavior beside an overall decline so that for very long ferromagnetic part ($L\gg d_F$), the mixing transmission diminishes. Nevertheless for reasonable lengths up to a micron we see $t^{\uparrow\downarrow}$ is still finite assuming $J\sim 10$meV which can be achieved using a magnetic insulator layer on top of graphene \cite{} or by Zeeman effect. This is a very special property of the spin pumping device we proposed here in contrast to conventional ferromagnetic spin pumping systems with large $J$ and very small $d_F$ so that the mixing transmission is negligible. As indicated in the inset of Fig. \ref{fig4}, the spin pumping itself oscillates with $L$ while the variation of doping $\mu_F$ only influences the amplitude of oscillations in spin current. 
\begin{figure}[tp]
 \includegraphics[width=0.9 \linewidth]{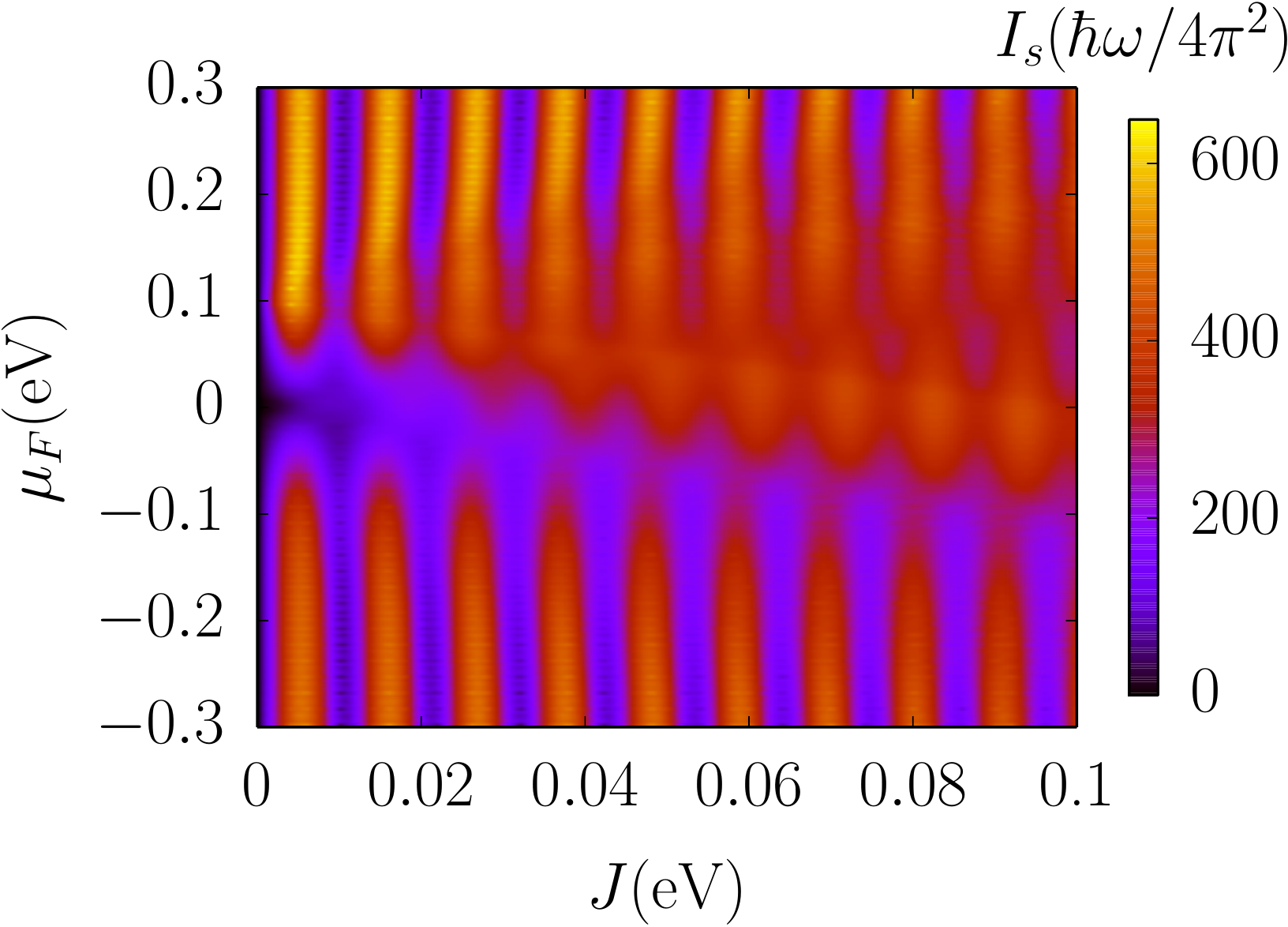}
  \caption{\label{fig5} Contour plot of spin pumping $I_{s,N1}^{\rm pump}$ as a function of exchange field strength $J$ and the doping $\mu_F$. The pumped spin is an oscillatory function of $J$ with the period of $\pi L/\hbar v_F$. The amplitude of spin current varies with $\mu_F$ and also a phase shift takes place in the oscillation over the region defined by $|\mu_F|<J$. Other parameters are chosen as $W=2 \mu m$, $\mu_{N1}=\mu_{N2}=0.2$eV and $ L=100$nm}
\end{figure}
\par
The behavior discussed above can also been observed from Fig. \ref{fig5} more clearly where we have contour plot revealing the dependence of the spin pumping on the spin splitting $J$ and the doping $\mu_F$. It shows that although the oscillations has the certain period of $\Delta J\sim 2\pi \hbar v_F/L $ however they suffer from a phase shift inside a certain region of the two dimensional plane of parameters $J,\mu_F$. This region can be defined more precisely by $|\mu_F|<J$ in which the carriers type inside the ferromagnet differs for two spin subbands, but outside this region both up and down spins are from either conduction or valence band. As a result when $|\mu_F|<J$ the minority (majority) spins feel an $npn$ ($nn'n$) barrier while for larger dopings $|\mu_F|>J$ both spin carriers feel the same type of the barriers during transmission through the ferromagnetic part. This influences the phases in the electrons transmissions and reflections due the chiral characteristics of electrons in graphene, which leads to an almost $\pi$ shift in the oscillations inside $|\mu_F|<J$ with respect to the regions outside. Due to this $\pi$ shift, depending on the values of spin splitting $J$, the spin pumping current can show different behaviors as a function of doping $\mu_F$. For lower values of $J$ the doping assists the spin pumping however at higher $J$ depending on the exact value the spin pumping can be decreased or increased as a function of the doping of magnetic region.  
At the end it should be mentioned that similar to N|F system, the amount of pumped spin differs upon changing the sign of the doping $\mu_F$. In fact when the ferromagnetic region's doping is the same as two N parts the spin pumping is more efficient than the situations where the doping of the ferromagnet is different.
\par
\begin{figure}[t]
 \includegraphics[width=0.53 \linewidth]{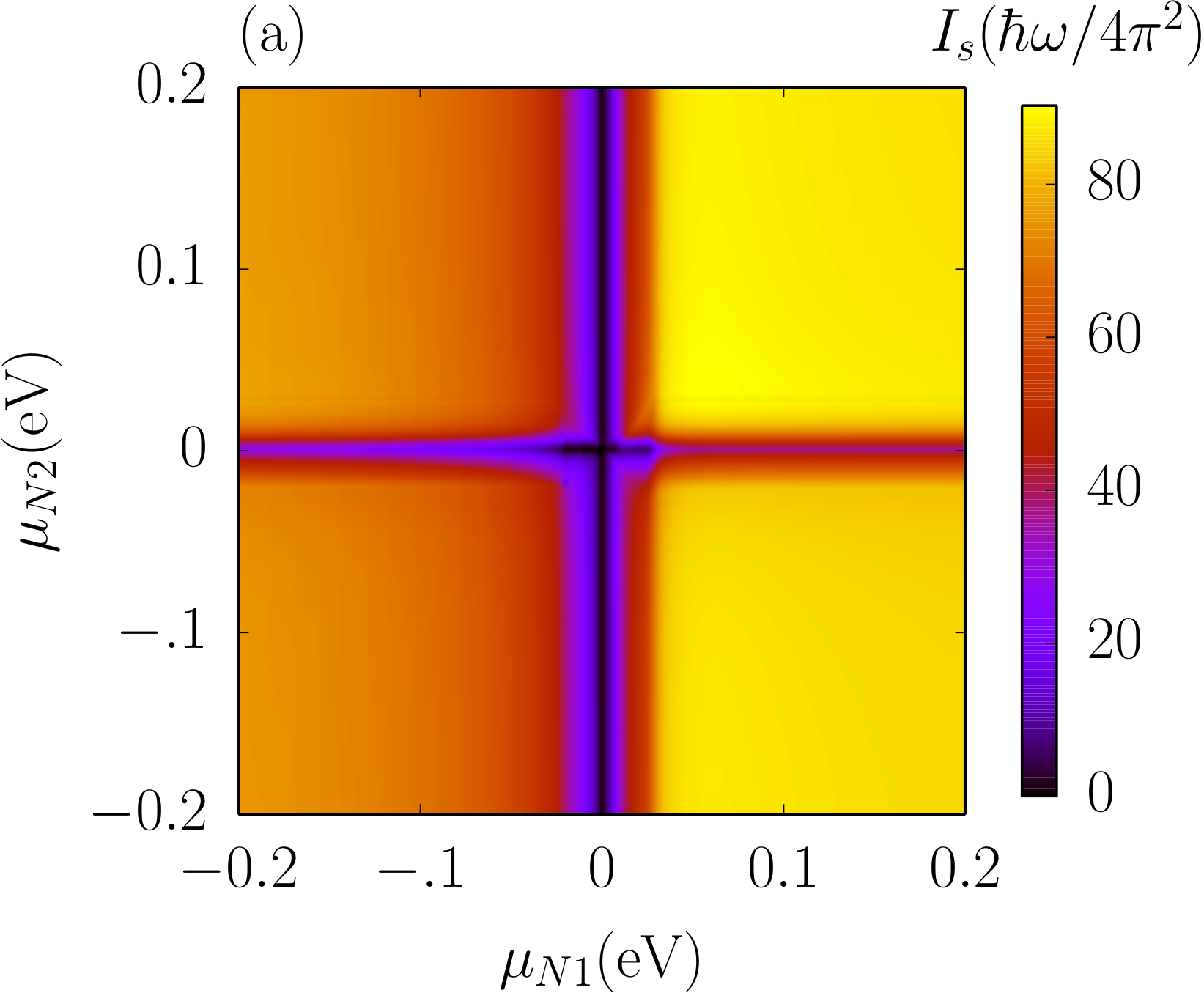}
  \includegraphics[width=0.45 \linewidth]{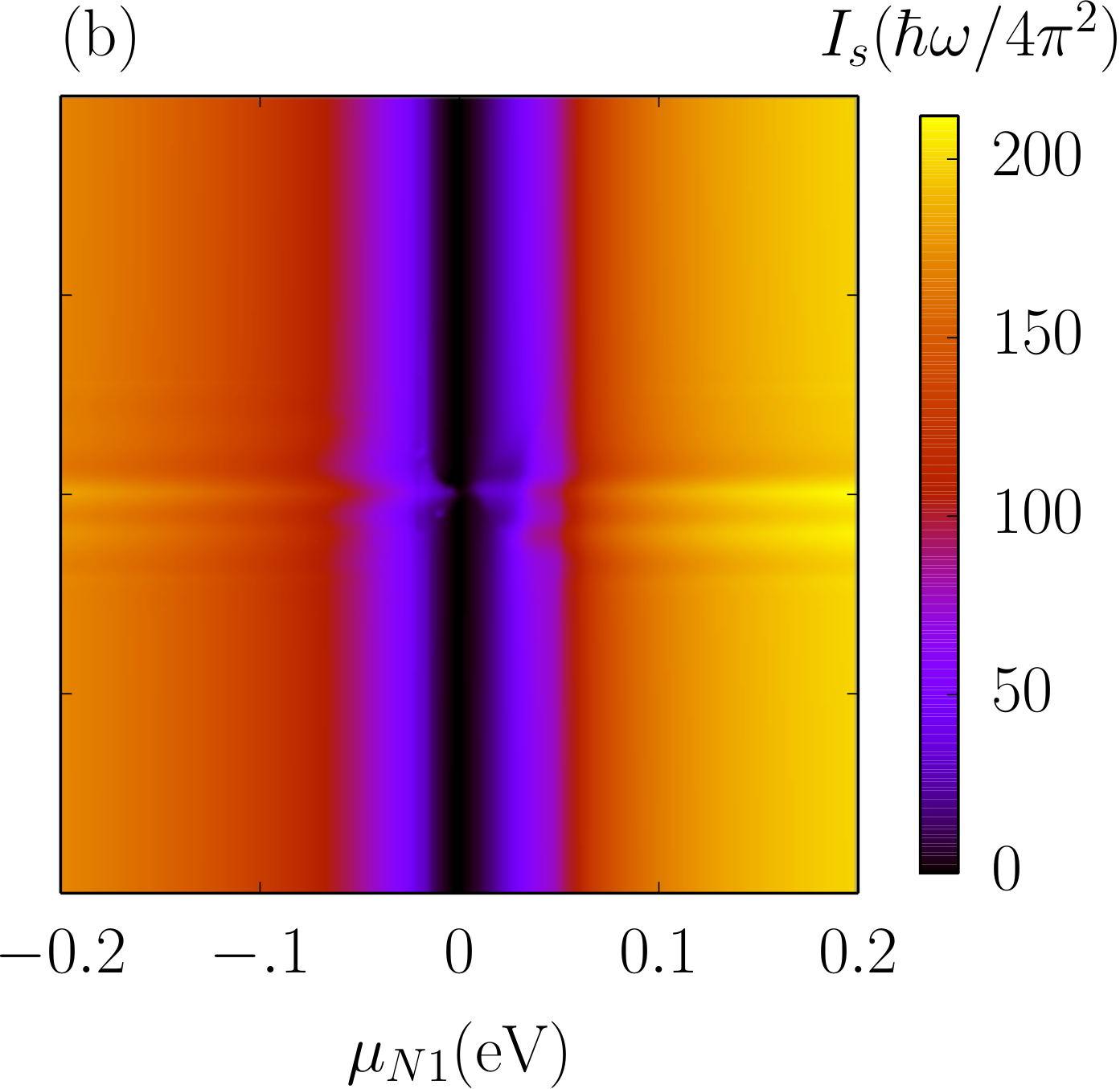}
  \caption{\label{fig6} Pumped spin current as a function of gate voltages on $N1$ and $N2$ regions for (a) $J=10$meV, (b) $J=40$meV. Other parameter are chosen as $\mu_{F}=20$meV, $L=50$nm, $W=1 \mu m$.}
\end{figure}

It is mentioned couples of times that spin pumping can be easily controlled by gate voltage in graphene. Therefore one may expect to have different amount of spin currents pumped into two normal regions in the N|F|N structure. Figure \ref{fig6} shows the variation of pumped spin to the first normal lead ($I_{s,N1}$) with the doping of two N electrodes for two different $J=10,40$meV. In both cases we see that except at the vicinity of low doping $\mu_{N2}$, $I_{s,N1}$ only depends on the doping of N1 electrode. This means varying the doping of any of electrodes has no effect on the spin pumping to the other N electrode, at higher dopings $\mu_{N_2}\gtrsim ||\mu_F|-J|$. At this regime the spin pumping into the two different N electrodes are almost independent of each other and can be controlled, separately. Then the dependence of spin current pumped into each N region on its doping reveals almost similar behavior as F|N system. So by increasing the doping $|\mu_{N1}|$, the corresponding $I_{s,N1}$ also increases and there is a region around $\mu_{N1}=0$ which due to the absence of spin filtering the spin pumping is suppressed. On the other hand when the doping of the second normal lead is low, it influences the spin pumping to the first one which depending on the value of $J$ (or equivalently $L$), it's effect can be constructive or destructive. We see that when $\mu_{N2}\sim 0$, the spin pumping $I_{s,N1}$ is absent for $J=10$meV (Fig. \ref{fig6}a), while for $J=40$meV (Fig. \ref{fig6}b) the pumped spin current is even enhanced. Intriguingly this reduction or enhancement of spin pumping caused by the doping of second lead is almost independent of doping amount of the first electrode. 
\par
An intriguing feature of results in Fig. \ref{fig6} is their asymmetry with respect to interchange of $\mu_{N1}$ and $\mu_{N2}$ which is very profound when one of the dopings is very small in the right panel with $J=40$meV. This asymmetry can be interpreted as difference in the spin current pumped into two different normal leads, since the amount of $I_{s, N1}$ when we interchange the dopings is exactly the same as $I_{s, N2}$ with original dopings. Therefore in our N|F|N device by tuning the dopings to be different inside two normal electrodes the so-called \emph{spin battery} effect can be arisen. Moreover by fine tuning the length of the ferromagnet or the spin splitting $J$, we can reach the regime in which low doping of N2 leads to enhancement in $I_{s,N1}$ and vice versa. Subsequently we can have maximum spin battery effect by choosing right values of $J$ and $L$ when one of the dopings is very low.
\par
It worths to comment on the possible experimental realization of the results presented here. We have investigated the effect of different doping in different parts as well as the roles of spin splitting strength and the geometry on spin pumping by rotating magnetization which needs to be verified in experiments.
Already spin injection and spin valve effects have been observed in graphene based devices. In addition, very recently, dynamical spin pumping has been reported by Tang \emph{et al.} \cite{Tang2013} which suggest the devices we have studied here can be implemented in real experiments. Here we have considered systems with widths and lengths up to micron size which lead to spin currents as large as thousands of $\hbar\omega$ with $\omega$ indicating the frequency of magnetization precession. Since we are dealing with metallic system the pumped spin current could be measured via inverse spin Hall effect (ISHE) which results in charge currents up to $I_{\rm ISHE}\sim 10^3 e\omega$. This means assuming the experimentally available gigahertz regime ($\omega\sim10^9s^{-1}$) the resulted charge current due to ISHE can reach $\sim 0.1\mu$A which can be easily measured. 
\par
Finally it must be emphasized that we focus on ballistic transport regime and some of the results we predict especially those related to quantum interference effects may only be seen in very clean samples. Nevertheless we believe some of the features related to the energy scales like doping and spin splitting are very general irrespective of transport regime. In particular even in the diffusive regime one must see qualitatively the same behavior of spin pumping as presented in Figs. \ref{fig1} and \ref{fig2}, since the proposed spin filtering mechanism only depends on the band structure properties.

\section{Conclusion}
In conclusion, we investigate the spin pumping in  graphene-based ballistic normal/ferromagnet (N/F) contacts via the rotating magnetization induced inside F region. We have employed the scattering theory framework for spin pumping to obtain the DC component of adiabatically pumped spin current. It is shown that the spin current can be suppressed or become significant as a function of doping in N and F regions. In particular the maximum spin current is achieved for $|\mu_F|=J$ when density of states for one of the spin species vanishes at the Dirac point. We relate this effect to the so called full spin filtering which enhances spin pumping in contrast to other transport characteristics like conductance of the N|F system. Interestingly, because of small induced spin splitting in graphene, the magnetic coherence length is quite large in contrast to metallic ferromagnets. As a result mixing transmission cannot be ignored which influences the spin pumping in hybrid N|F|N structure. Then the spin current shows an oscillatory behavior versus the length $L$ with a period of $d_F=\pi\hbar v_F/J$ because of quantum interferences. Moreover we reveal that the spin current pumped into two normal sides can be different and controllable with their doping which leads to the spin battery effect with tunable polarization and power.

\bibliography{grph}

\end{document}